\documentclass[sigconf]{acmart}

\usepackage[english]{babel}
\usepackage{blindtext}
\renewcommand\footnotetextcopyrightpermission[1]{} 
\setcopyright{none}
\usepackage{graphicx}
\usepackage{subcaption}

\acmDOI{}

\acmISBN{}

\acmPrice{}

\begin{document}
\title{Programmable Packet Scheduling with Dynamic Reordering at Line Rate}
\acmConference[SIGCOMM’26]{ACM SIGCOMM 2026}{August 17--21, 2026}{Denver, Colorado, USA}
\acmYear{2026}
\copyrightyear{2026}


\author{Zekun Wang, Binghao Yue, Yichen Deng, Weitao Pan, Jiangyi Shi, Yue Hao}
 \affiliation{
	   \institution{Xidian University}
	 }

\renewcommand{\shortauthors}{Zekun Wang.et al.}

\begin{abstract}
High-speed switch packet scheduling demands both line-rate performance and programmability. Existing programmable hardware scheduling models, such as PIFO and PIEO, can express a broad range of scheduling algorithms; however, their semantics are restricted to packet-level ordering and cannot dynamically reorder buffered packets, which limits the support for dynamic-ordering algorithms such as pFabric.

To overcome this limitation, we propose UIFO (Update-In-First-Out), a new programmable scheduling model that introduces a two-level abstraction over classes and packets. UIFO enables dynamic updates to the scheduling order at the class level while preserving in-order packet scheduling within each class, thereby supporting dynamic reordering of already-buffered packets. Furthermore, UIFO remains fully compatible with and generalizes existing PIFO and PIEO models.

We implement a hardware prototype of UIFO based on priority-queue designs and evaluate it on an FPGA platform and in a 28 nm ASIC process. Overall, UIFO significantly enhances scheduling expressiveness and maintains favorable scalability while sustaining 100 Gbps line-rate throughput.
\end{abstract}


\maketitle

\section{Introduction}
Packet scheduling governs the timing and ordering of packet transmissions at a network device’s egress, directly affecting fairness, latency, and quality of service (QoS). Its fundamental objective is to determine when and in what order packets queued at an output port should be transmitted \cite{pifo}. As cloud computing, large-scale datacenters, and latency-sensitive applications continue to evolve, network traffic has become increasingly diverse and dynamic, making it difficult for any single scheduling algorithm to meet all operational requirements. In practice, network operators rely on a variety of scheduling policies, each optimized for specific objectives—for example, pFabric \cite{pfabric} aims to minimize flow completion time(FCT), Priority Flow Control (PFC) \cite{pfc} targets lossless transport, token-bucket mechanisms are widely used for traffic shaping \cite{tb}, and fairness-oriented schedulers such as WFQ \cite{wfq} and DRR \cite{drr} seek bandwidth-proportional allocation.
\\With the advent of SDN, the flexibility of the control plane has been substantially improved; however, similar flexibility is also needed in the data plane, particularly for queue management and scheduling \cite{no_silver}. Since no universally optimal packet scheduling algorithm exists in practice \cite{ups}, modern networks increasingly demand programmable schedulers that can support a wide range of policies at line rate. Unfortunately, existing programmable hardware scheduling models largely confine scheduling decisions to packet-level ordering. Once a packet is enqueued, its relative scheduling position typically cannot be updated in response to subsequent state changes. This limitation makes it difficult to naturally express many algorithms that require runtime information to dynamically reorder buffered packets, while still preserving line-rate performance.
\\To enable programmability across a broad range of scheduling algorithms and to improve the flexibility of hardware-programmable schedulers, recent years have seen the emergence of programmable scheduling models represented by PIFO \cite{pifo} and PIEO \cite{pieo}, which largely define the capability boundary of today’s hardware-programmable packet scheduling. While these models are expressive enough to realize most scheduling algorithms, they share a fundamental limitation: once a packet is enqueued, its relative scheduling order cannot be dynamically updated in response to subsequently arriving packets or changes in system state.
For instance, pFabric aims to minimize FCT by always prioritizing packets from the flow with the smallest remaining size. When a flow with a smaller remaining size arrives, the algorithm semantically requires that all buffered packets belonging to this flow be advanced in the service order. However, under existing PIFO and PIEO abstractions, enqueuing a new packet alone cannot alter the relative order among multiple packets already buffered in the queue. Consequently, such algorithms are difficult to implement efficiently in programmable hardware schedulers.
From an abstraction perspective, the key difference between PIFO and PIEO lies in how the set of schedulable packets is constructed. Nevertheless, both models ultimately adhere to the same scheduling semantics: performing packet-level ordering within a given candidate set. This packet-centric view makes it hard for state updates to directly affect the service order of buffered packets, thereby becoming a fundamental bottleneck for dynamically adaptive scheduling policies.
\\A large body of prior work on programmable packet schedulers has primarily focused on improving implementation efficiency—such as queue scalability and throughput—within the aforementioned scheduling models. However, there is still a lack of a direct and general abstraction that supports updating the service order of already-buffered packets. To bridge this gap, we propose a new programmable scheduling model, UIFO (Update-In-First-Out).
UIFO introduces a two-level scheduling semantics. The scheduler first selects among classes and allows the priority of a class to be updated based on runtime state; it then schedules packets within the chosen class according to packet priorities. This hierarchical design enables UIFO to preserve fine-grained packet-level scheduling, while allowing flow-/queue-level state updates to immediately reshape the service order of buffered packets.
Built on this model, UIFO can naturally express dynamic-ordering algorithms such as pFabric, while remaining compatible with existing programmable scheduling abstractions including PIFO and PIEO. We implement UIFO using an update-enabled priority queue and a multi-priority-queue group, and evaluate its resource cost and timing characteristics on a Xilinx XCVU13P FPGA as well as in a 28nm ASIC process. We will open-source the UIFO implementation in camera ready version.\\
In this paper, we make the following contributions:
\begin{itemize}
	\item Identify a fundamental semantic limitation shared by existing programmable scheduling abstractions, including PIFO and PIEO: scheduling decisions are ultimately applied at the packet level, preventing state updates from directly reordering buffered packet.
	\item Proposed UIFO, a new programmable scheduling abstraction that elevates subsets of data packets to first-class scheduling objects, enables reordering of data packets through class-based scheduling, while remaining fully compatible with previous models.
	\item Demonstrate that UIFO can naturally express a broad range of scheduling algorithms---including dynamic, flow- or queue-centric policies such as pFabric,\allowbreak PFC---that are difficult or inefficient to realize under existing abstractions.
	\item Implemented UIFO in hardware using a priority queue that supports update operations and a multi-priority-queue group, and evaluated its resource cost and timing on FPGA and ASIC platforms, showing that the increased expressive power comes with a moderate hardware overhead.
\end{itemize}
\textbf{This work does not raise any ethical issues.}
\section{Background}
\subsection{Packet Scheduling Algorithms}
Packet scheduling algorithms are a key mechanism in network devices for determining the transmission order and timing of competing packets on an output link. Their goal is to provide differentiated guarantees across traffic flows—such as latency, throughput, fairness, and QoS—under constrained link bandwidth and buffer resources. Depending on the different classification criteria, packet scheduling algorithms can be further categorized.
\subsubsection{Classification by Link Utilization}
All packet scheduling algorithms make two fundamental decisions: the service order of packets and their transmission time. Based on how a scheduler behaves when the output link is idle, scheduling algorithms are typically classified into work-conserving and non-work-conserving algorithms \cite{pifo}.Work-conserving algorithms focus primarily on determining packet transmission order: whenever the link becomes idle, the scheduler always transmits the packet at the head of the queue, ensuring that the output link never remains idle as long as there are buffered packets. In contrast, non-work-conserving algorithms emphasize transmitting packets at the correct time. If buffered packets do not satisfy the required transmission conditions (i.e., the eligibility predicate evaluates to false), the scheduler defers transmission even when the link is idle.
\subsubsection{Classification by Dynamic Ordering}
Depending on whether the scheduling order of buffered packets may change upon the arrival of new packets, scheduling algorithms can be classified into static-ordering and dynamic-ordering schemes. In static-ordering algorithms, once a packet enters the queue, its priority remains unchanged, as in LSTF \cite{lstf}, WFQ \cite{wfq}, DRR \cite{drr}, and STFQ \cite{stfq}. Such algorithms can be mapped straightforwardly onto conventional priority-queue-based models. In contrast, dynamic-ordering algorithms exhibit the following characteristic: the service order of buffered packets may be revised as new packets arrive. For example, in pFabric, when an newly arrived packet belongs to a flow with a shorter remaining size, the scheduler must transmit ahead the already-buffered packets of that flow, thereby reordering multiple packets in the queue. Similar dynamic reordering strategies include D3 \cite{d3}, DDRR \cite{ddrr}, and DBA \cite{dba}. Although existing hardware scheduling models can precisely support static-ordering algorithms, they remain notably limited in both expressing and efficiently implementing dynamic-ordering algorithms.
\subsection{Hardware Programmable Scheduling Model}
Currently, programmable scheduling models that have been implemented in hardware and extensively studied mainly fall into two categories: PIFO \cite{pifo} and PIEO \cite{pieo}.
\subsubsection{Push-In-First-Out}
In the PIFO scheduling programming model, users provide a small program to compute a packet’s rank, which determines its priority. Combined with a single priority queue, PIFO can implement scheduling algorithms that specify either transmission timing or transmission order. However, its key limitation is that the service order of already-buffered packets cannot be changed in response to subsequently arriving packets.
In recent years, a large body of work has focused on improving the scalability and throughput of hardware implementations of the PIFO model, proposing both exact and approximate designs. Among them, BBQ \cite{bbq} is a representative exact implementation: it uses a tree-based structure to index priorities and supports logical partitioning by segmenting the priority field.
\subsubsection{Push-In-Extract-Out}
The PIEO scheduling model can be viewed as a generalized extension of PIFO. Its key idea is to abstract scheduling as selecting, at any time, the highest-priority eligible element (i.e., the smallest-ranked eligible element). By introducing an eligibility predicate, PIEO can capture both scheduling order and transmission time, thereby supporting a broader class of scheduling algorithms. The model exposes three primitive operations—\(enqueue(f)\), \(dequeue()\), and \(dequeue(f)\)—which respectively support enqueuing, dequeuing from the head, and dequeuing an element from an arbitrary position.
\subsection{Limitation}
\begin{figure}[t]
	\centering
	\includegraphics[width=0.98\columnwidth]{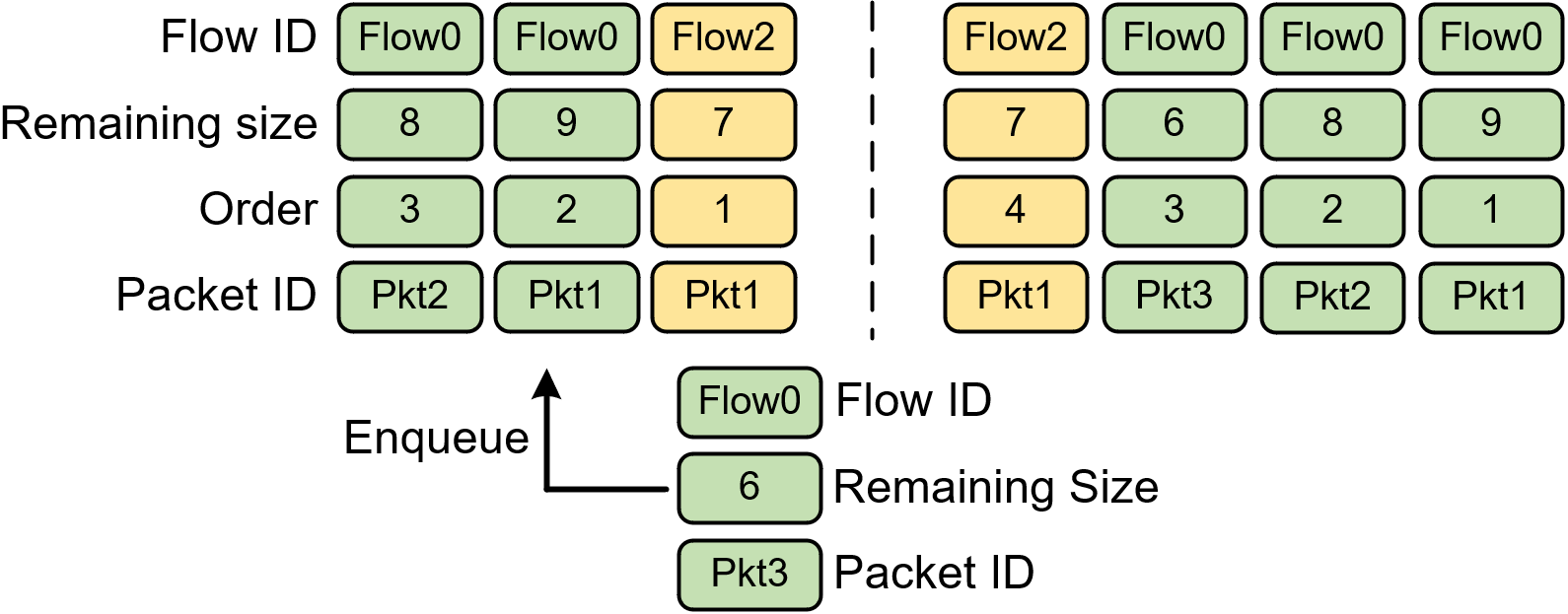}
	\caption{Mechanism of the pFabric Algorithm.}
	\label{fig:1}
\end{figure}

Although both PIFO and PIEO have been realized in hardware, they remain fundamentally limited in expressing dynamic-ordering scheduling algorithms. We illustrate this limitation using pFabric, a dynamic algorithm that achieves near-optimal FCT. pFabric always prioritizes packets from the flow with the smallest remaining flow size, while enforcing first-in-first-out(FIFO) order within each flow.
\\Fig.~\ref{fig:1} depicts a representative scenario: when \(flow0.pkt3\) arrives, its associated remaining flow size becomes the smallest among all flows currently buffered in the queue. Consequently, the scheduler should advance the already-buffered packets of this flow and transmit them earlier. However, since PIFO does not support modifying the relative order of buffered packets, it cannot realize this scheduling behavior.
\\Although PIEO can, in principle, update the scheduling order of buffered packets by dequeueing a specific element and re-enqueueing it, this approach incurs substantial overhead in practice. Consider the three buffered packets in Fig.~\ref{fig:1}. Since PIEO requires 4 clock cycles to complete a single operation, updating the position of one packet would take at least 8 cycles. As a result, enqueuing one new packet while reordering the other three buffered packets would require 28 clock cycles in total. This overhead grows further with the number of buffered packets, making it difficult to meet performance requirements in high-throughput settings.
Moreover, to guarantee \(O(1)\) time complexity for all operations, PIEO implements a queue of \(N\) elements using \(2\times\surd(N)\) sublists of size \(\surd(N)\), resulting in a storage footprint of \(2\times N\) elements and thus low space efficiency. DR-PIFO \cite{dr_pifo} attempts to extend PIFO’s expressiveness via dynamic reordering and forced-dequeue mechanisms; however, it requires supporting deletion from an arbitrary position in the priority queue and searching the buffer for the head packet of a target flow. To date, no practical hardware implementation has been demonstrated.
Similar challenges also arise in other dynamic-ordering algorithms. For example, in D3 \cite{d3}, end hosts dynamically adjust sending rates based on the remaining flow size and the remaining deadline. When link capacity becomes insufficient, D3 employs flow quenching to reduce the scheduling priority of packets from selected already-buffered flows, which likewise requires the scheduler to dynamically reorder buffered packets. These limitations motivate a new abstraction that can efficiently update the service order of buffered packets.
\begin{figure*}[t]
	\centering
	\includegraphics[width=1.98\columnwidth]{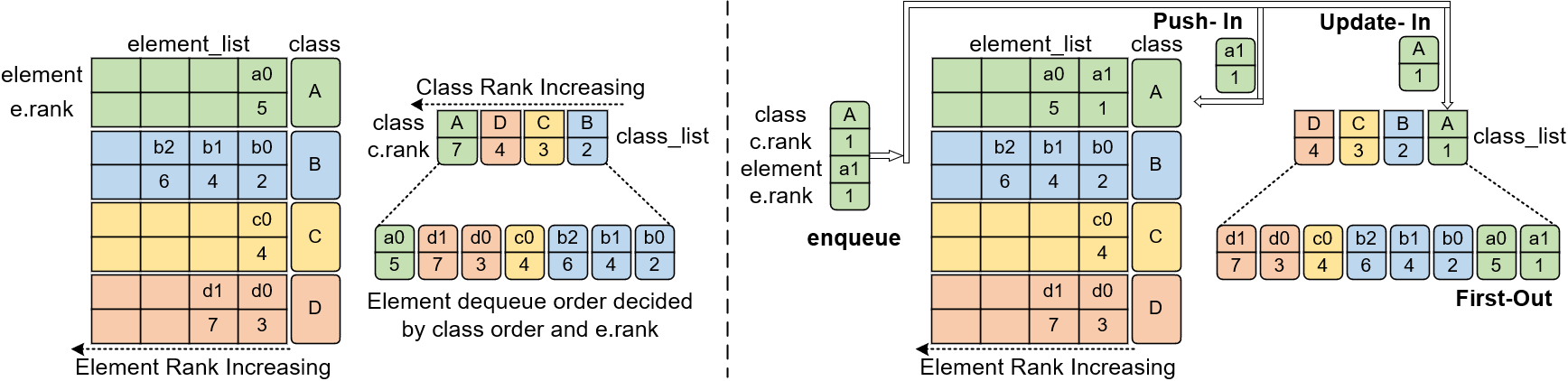}
	\caption{UIFO Programmable Model}
	\label{fig:2}
\end{figure*}
\section{Update-In-First-Out(UIFO)}
In this section, we describe the design motivation, programming framework, and compatibility of UIFO.
\subsection{Design Motivation}
From the perspectives of transmission time and service order, both PIFO and PIEO share a common abstraction: a scheduling algorithm operates directly on a sequence of packet-granularity elements, where each decision selects the single highest-priority element from a given candidate set. This semantics is sufficient for describing many classical scheduling policies, but it exhibits a fundamental gap for a class of dynamic reordering algorithms. Such algorithms often make scheduling decisions at the granularity of flows or queues, rather than individual packets. Concretely, during runtime, the scheduler dynamically adjusts the service order across flows/queues based on flow-/queue-level state (e.g., estimated remaining flow size, congestion feedback). Crucially, these adjustments should take effect immediately on packets that are already buffered, instead of only influencing future arrivals. Models that only rank packet-level elements can easily degenerate this semantics into “only affecting newly enqueued packets,” thereby deviating from the intended algorithmic behavior.
\\We further observe that PIEO’s dequeue process can be abstracted as selecting and dequeuing the highest-priority element from the subset of elements that satisfy the eligibility predicate. Similarly, in PIFO designs that support logical partitioning \cite{bbq}, elements can be enqueued into a priority queue associated with a designated logical partition, while dequeue operations can be restricted to selecting the highest-priority element within a specific partition.
Since logical partitions likewise correspond to restricted candidate sets, the mechanisms above can be unified under the same abstraction: select the highest-priority element to serve from a specified set of elements. This unified view exposes the common structure of existing abstractions: they implicitly include a subset-selection step (where a subset is formed via an eligibility predicate or logical partitioning), followed by packet-level scheduling that selects the highest-priority packet within the chosen subset. The key problem, however, is that neither the subset-selection itself nor the relative service order across subsets is explicitly modeled or programmable. As a result, it is difficult to express dynamic ordering policies at the flow/queue granularity—especially those requiring reordering of already-buffered packets. In other words, while PIFO/PIEO can “assign packets to a subset,” they lack an abstraction for “how subsets should be scheduled.” Eligibility predicates and logical partitioning address classification, but they do not elevate subset scheduling into a first-class programmable object. Motivated by this observation, we propose adding an explicit layer of scheduling logic over element subsets, on top of the existing abstraction of “selecting the highest-priority element within a set.” This enables the scheduler to not only select packets within a subset, but also decide which subset to serve first. 
\subsection{UIFO Programmable Model}
This section presents the programmable scheduling model of UIFO. UIFO explicitly distinguishes inter-class scheduling and intra-class scheduling, decomposing scheduling logic into two layers. Accordingly, it introduces two types of schedulable objects—Class and Element—which correspond to inter-class and intra-class scheduling, respectively. The scheduler is built upon two priority queues that order these objects by their rank values. To ensure deterministic scheduling behavior, we require that the IDs of schedulable objects be unique within their respective priority queues. Each packet is represented as an Element, whose scheduling attributes are defined as follows:
\begin{itemize}
	\item \textbf{Element ID}, which uniquely identifies a packet;
	\item \textbf{Element Rank} ($e.rank$), the packet priority computed by the scheduling algorithm;
	\item \textbf{Class ID}, which indicates the Class to which the packet belongs;
	\item \textbf{Class Rank} ($c.rank$), which specifies the class-level priority of the packet's associated Class.
\end{itemize}
Fig.~\ref{fig:2} illustrates the programming framework of UIFO. UIFO consists of a two-level priority structure with two layers of semantics:
\begin{itemize}
	\item \textbf{Intra-class scheduling}. Each Class $c$ is associated with an \textit{element\_list}, which stores all Elements belonging to that class. The \textit{element\_list} is ordered by $e.rank$, enabling packet-level scheduling within the class.
	\item \textbf{Inter-class scheduling}. All Classes are managed by one unified \textit{class\_list}, which is ordered by $c.rank$ to control the service order among classes. Each class in \textit{class\_list} is uniquely associated with exactly one \textit{element\_list}.
\end{itemize}
Since each packet carries its associated \textit{Class ID} and the corresponding $c.rank$ upon enqueue, the arrival of a new packet belonging to an existing class inevitably requires updating that class's priority in \textit{class\_list}, thereby changing the scheduling order across classes. Such class-level priority updates are key to enabling UIFO's dynamic scheduling semantics. UIFO defines its scheduling behavior using two fundamental primitives.
\\\textbf{Enqueue.} The enqueue operation \(enqueue(e)\) inserts an element $e$ into the scheduling model. Each element $e$ carries its packet-level priority $e.rank$, as well as its associated class $c$ and the corresponding $c.rank$. The operation proceeds as follows:
\begin{enumerate}
	\item Element-level insertion: the element $e$ is inserted into the \textit{element\_list} of its associated class $c$ and order it by $e.rank$ (Push-In);
	\item Class-level update: if class $c$ already exists in \textit{class\_list}, update its scheduling order according to the new $c.rank$; otherwise, class $c$ is inserted into \textit{class\_list} with priority $c.rank$ (Update-In);
	\item If the class priority does not change, the relative ordering of classes remains unchanged (Hold).
\end{enumerate}
\textbf{Dequeue.} The dequeue operation $dequeue()$ follows the rules below:
\begin{enumerate}
	\item Select the highest-priority (head) class $c$ from \textit{class\_list} (First-Index);
	\item From the corresponding \textit{element\_list} of class $c$, dequeue the element $e$ with the highest packet-level priority (First-Out);
	\item If the \textit{element\_list} of class $c$ becomes empty after dequeuing, class $c$ is dequeuing from \textit{class\_list}; otherwise, the ordering of classes remains unchanged (First-Out).
\end{enumerate}
Therefore, at any time, UIFO follows the scheduling principle of serving the element with the highest packet-level priority from the class with the highest class-level priority.
\\Structurally, UIFO is a two-level scheduling data structure: the outer level schedules among classes, while the inner level performs packet-level scheduling within each class. Upon enqueue, an element insertion simultaneously triggers a class-priority update and a packet-level insertion, tightly coupling state updates with scheduling decisions. When the service order across classes changes, the effect is equivalent to reordering multiple \textit{element\_list}s, thereby dynamically reshaping the service order of buffered packets without explicitly relocating individual packets. This property allows UIFO to preserve fine-grained packet-level control (e.g., different packets within the same flow may still have different priorities), while enabling flow-/queue-level dynamic decisions to take effect immediately. As a result, UIFO naturally expresses dynamic-ordering algorithms that depend on flow-/queue-level state updates. A detailed analysis of UIFO's expressiveness for additional algorithms is provided in Section~4.
\subsection{Compatibility}
This section discusses the relationship between UIFO and two representative programmable scheduling models, namely PIFO and PIEO.
\\PIFO can be viewed as a single-level programmable scheduling model whose core data structure is a queue that maintains a total order over elements according to their priorities. The scheduler always dequeues the globally highest-priority element. When logical partitioning is not considered, PIFO's scheduling behavior is fully determined by the packet priority, which can be mapped onto UIFO's two-level scheduling structure via \emph{priority field segmentation}. Specifically, the priority field of each element in PIFO is split into a high-order field and a low-order field:
\begin{itemize}
	\item The high-order field determines the class of an element and serves as the $c.rank$ of that class, which is used to order classes in \textit{class\_list};
	\item The low-order field is used as the packet-level priority, which orders elements within the corresponding \textit{element\_list}.
\end{itemize}
\begin{table*}[t]
	\centering
	\caption{Comparison of Scheduling Model Abstractions}
	\label{tab:1}
	\setlength{\tabcolsep}{2pt} 
	\resizebox{0.9\textwidth}{!}{ 
		\begin{tabular}{c c c c c c} 
			\toprule
			Model & Scheduling Object & Element Classification & Outer Scheduling Object & Inner Packet Selection Rule & Supports Dynamic Reordering \\
			\midrule
			PIFO\cite{pifo} & Packet & No explicit classification & None & By per-packet priority & No \\
			Logical Partition PIFO\cite{bbq} & Packet & Logical partition & Logical partition & By per-packet priority & No \\
			PIEO\cite{pieo} & Packet & Eligibility predicate & Eligibility set & By per-packet priority & No \\
			UIFO & Class + Packet & Programmable class & Explicit class & Per-packet priority within class & Yes \\
			\bottomrule
		\end{tabular}
	}
\end{table*}
Under this mapping, UIFO's class-level scheduling and packet-level scheduling together form an equivalent total ordering: the head class in \textit{class\_list} necessarily corresponds to the set of elements with the current globally highest priority, and the head element in that class's \textit{element\_list} is exactly the highest-priority element in the entire system. Therefore, under this configuration, UIFO's scheduling decisions are strictly identical to those of the original PIFO. 
\\When PIFO enables logical partitioning, its scheduling behavior can be described as follows: upon enqueue, an element is inserted into the priority queue of a designated partition; upon dequeue, the scheduler selects the highest-priority element within a specified partition. This mechanism can also be naturally mapped to UIFO. Specifically, each logical partition is mapped to a class; all classes are ordered in \textit{class\_list} according to FIFO. The per-partition priority queue corresponds to UIFO's \textit{element\_list}, which is ordered by packet-level priority. To dequeue from a designated logical partition, it suffices to temporarily elevate the priority of the corresponding class to the highest value, thereby ensuring that the scheduler selects and dequeues the highest-priority element within that class. As a result, under this mapping, UIFO fully reproduces PIFO's scheduling semantics in the logical partitioning function.
\\In the PIEO model, the eligibility predicate is typically defined by \textit{send time}: a packet is eligible for scheduling only if its send time is no later than the current time. UIFO can express PIEO equivalently by using send time as the criterion for class construction. Specifically, packets with the same send time—or whose send times fall within the same time interval—are mapped to the same class, whose priority is determined by the send time. Upon enqueue, each packet is assigned to the \textit{element\_list} of the corresponding class based on its send time. At any scheduling instant, the head class in \textit{class\_list} necessarily corresponds to the group of packets with the earliest send time. When the send time of this class satisfies the eligibility condition, all packets in its \textit{element\_list} become eligible for scheduling. Since the \textit{element\_list} is ordered by packet-level priority, the packet selected from this class is exactly the highest-priority packet among all eligible packets. Therefore, under this configuration, UIFO’s scheduling behavior is consistent with the scheduling semantics of PIEO.
\\It is worth noting that, in PIEO, update operations are typically performed at the granularity of individual packets. A packet’s send time or priority is updated by \emph{dequeue(f) and enqueue(f)} opeartions, thereby modifying the scheduling order. In contrast, UIFO performs updates at the granularity of a class, which may consist of multiple packets. A single update to a class’s priority can simultaneously affect the service order of all buffered packets belonging to that class. This class-granular update mechanism preserves the scheduling semantics of PIEO, while significantly reducing the number of update operations, thereby lowering implementation complexity and improving scheduling efficiency.
\\To facilitate understanding the relationship between PIFO, PIEO, their variants, and UIFO from a unified perspective, Table~\ref{tab:1} summarizes a comparative abstraction of each model. As shown, despite differences in classification mechanisms between PIFO, PIEO, and logical partition PIFO, their core functionality revolves around performing packet-level ordering within a set. In contrast, UIFO explicitly models the set as a schedulable class, thus introducing programmability for class-level scheduling order.
\\In summary, as shown in Table~\ref{tab:1}, PIFO, PIEO, and PIFO with logical partitioning can be abstracted as performing packet-level priority selection within a given set of elements, with their primary differences lying in how the candidate set is constructed. 
UIFO explicitly models the element set as schedulable classes and modifies the service order across sets by updating class-priority, thereby introducing the capability of reordering buffered data packets while maintaining compatibility with packet-level scheduling semantics—a function not natively supported by PIFO and PIEO. 
Furthermore, when implementing flow-level scheduling equivalent to that of PIFO/PIEO, UIFO can model elements as the head packets of each flow queue to reproduce their flow-scheduling behaviors. When expressing algorithms that rely on flow state updates, UIFO can also model elements as intra-flow packets and reflect changes in flow states through class-level priority updates, thereby achieving dynamic adjustment of the inter-flow scheduling order. Therefore, the scheduling semantics of PIFO and PIEO can both be implemented by UIFO under specific configurations, and UIFO provides enhanced runtime update capabilities.
\section{The Expressiveness of UIFO}
In this section, we express a variety of dynamic-ordering and static-ordering algorithms using the UIFO model. Leveraging the two primitive operations introduced in Section~3.2, we show how UIFO can be used to implement these algorithms.
\subsection{pFabric}
pFabric~\cite{pfabric} decouples flow scheduling from rate control and dynamically adjusts flow scheduling priorities to minimize average FCT. It is a work-conserving, runtime-state-driven dynamic-ordering algorithm whose scheduling decisions are naturally made at the granularity of flows, rather than individual packets. UIFO can express pFabric's scheduling semantics in a direct and natural manner. 
\begin{verbatim}
enqueue(e):{
    c = Class(flow)
    c.rank = remaining_size(flow)
    class_list.enqueue(c)     # update c.rank
    c.element_list.enqueue(e)
}
dequeue():{
    c = class_list.head
    e = c.element_list.dequeue()
    if(c.element_list empty):
        class_list.dequeue()
}
\end{verbatim}
Specifically, we define class at the granularity of flows: all packets belonging to the same flow are mapped to the same class, and element $e$ is modeled as an intra-flow data packet. The $c.rank$ is determined by the remaining size of the flow, i.e., $c.rank = \mathrm{remaining\_size}(\textit{flow})$. For intra-class scheduling, pFabric does not impose packet-level prioritization; therefore, the \textit{element\_list} within each class follows FIFO order.
\\When a packet arrives, it carries an updated (smaller) remaining flow size. The $enqueue(e)$ operation triggers an Update-In on the corresponding class in \textit{class\_list}, thereby dynamically adjusting the scheduling order among flows. As a result, the class at the head of \textit{class\_list} always represents the flow with the smallest remaining size. During dequeue, UIFO’s $dequeue()$ operation first selects the class with the smallest class rank (i.e., the highest priority), and then dequeues a packet from the corresponding \textit{element\_list}. Consequently, at each scheduling decision, UIFO guarantees that the selected packet belongs to the flow with the smallest remaining size among all buffered flows, thereby strictly adhering to pFabric’s scheduling semantics.
\subsection{Priority-based Flow Control}
Priority-based Flow Control (PFC)~\cite{pfc} is a widely deployed lossless networking mechanism in datacenter networks. When a network device detects congestion, PFC can temporarily pause the transmission of selected priority queues, ensuring that higher-priority traffic is delivered in a timely manner. PFC pause frames explicitly carry both the priority queues to be paused and the duration of the pause. As a result, PFC’s scheduling behavior depends on runtime state updates and allows the output link to remain idle during pause intervals, classifying it as a non-work-conserving dynamic-ordering scheduling algorithm.
\\UIFO can express PFC in a natural and direct manner. Specifically, each priority queue is mapped to a class, and packets belonging to the same queue are assigned to the same class, with element $e$ modeled as the head packet of each flow. Since packets within a queue do not require further prioritization, the corresponding \textit{element\_list} follows FIFO order for intra-class scheduling. The $c.rank$ is used to characterize the transmission time of the queue, thereby precisely capturing PFC’s pause semantics.
\begin{verbatim}
enqueue(e):{
    c = Class(queue)
    if (pause queue_id):
        c.rank = current.time+pause.time
        class_list.enqueue(c)   # update c.rank
    else:
        c.rank = send.time
        class_list.enqueue(c)   # hold c.rank
        e.rank = 1
        c.element_list.enqueue(e) 
}
dequeue():{
    c = class_list.head
    if(c.rank <= current.time):
        e = c.element_list.dequeue()
        if(c.element_list empty):
            class_list.dequeue() 
}
\end{verbatim}
Upon receiving a PFC pause frame targeting a specific queue, the scheduler computes a new send time for that queue based on the current time and the pause duration, thereby deferring its transmission. The priority update is realized by performing an \textit{enqueue(e)} operation on the corresponding class in \textit{class\_list}. Since PFC pause frames are control frames rather than data packets, updating the class priority does not require inserting any element into the associated \textit{element\_list}.
\\Under normal operation, the queue’s scheduled send time remains unchanged. When data packets arrive, they are mapped to the corresponding class based on their queue and inserted into the associated \textit{element\_list}. During dequeue, the scheduler simply checks whether the $c.\mathrm{rank}$ of the head class in \textit{class\_list} satisfies the send-time condition. If the class rank is no greater than the current time, one packet is dequeued from that class’s \textit{element\_list}; otherwise, no packet is transmitted. Through this mechanism, UIFO precisely defers the transmission of all packets in the specified queue until the send-time condition is satisfied.
\subsection{Worst-case Fair Weighted Fair Queuing}
WF2Q+ (Worst-case Fair Weighted Fair Queuing)~\cite{wf2q} is a classic fairness-oriented scheduling algorithm. During scheduling, WF2Q+ maintains a global virtual time and, at any scheduling instant, selects for transmission the packet with the smallest Virtual Finish Time (VFT) from the subset of packets whose Virtual Start Time (VST) satisfies $\mathrm{VST} \leq$ current virtual time. Since there may exist instants at which no packet satisfies this condition, WF2Q+ is a non-work-conserving scheduler. For each packet, both VST and VFT are determined at enqueue time and remain unchanged throughout its residence in the queue; thus, WF2Q+ can also be regarded as a static-ordering scheduling algorithm.
\\WF2Q+ can be naturally mapped onto the UIFO programming model. We use VST as the criterion for class construction, grouping packets with the same VST into the same class, where element $e$ is modeled as the head packet of each flow. The $c.rank$ is directly determined by the corresponding VST, while packets within a class are ordered by their VFT, thereby realizing the scheduling order required by WF2Q+. The computation of VST and VFT follows the same method used in the PIEO model~\cite{pieo}.
\begin{verbatim}
enqueue(e):{
    e.VST = calculate_VST(e)
    c = Class(VST)
    c.rank = e.VST
    class_list.enqueue(c) # update c.rank
    e.rank = calculate_VFT(e)
    c.element_list.enqueue(e) 
}
dequeue(): same with PFC's dequeue function.
\end{verbatim}
Upon enqueue, the packet’s VST is used to determine its corresponding class and is set to the $c.rank$ of that class. The packet is then inserted into the \textit{element\_list} of that class, where it is ordered by its VFT. During dequeue, the scheduler first selects the class with the smallest $c.rank$. When the VST of this class is less than or equal to the current virtual time, the packet with the smallest VFT is dequeued from the \textit{element\_list}.
Through this mechanism, UIFO ensures that packets are always selected from the set with the smallest VST, and within that set, the packet with the earliest VFT is chosen for transmission. This guarantees that the scheduling behavior strictly follows the WF2Q+ semantics.
\\It is important to note that, unlike dynamic ordering algorithms such as pFabric or PFC, which depend on runtime state updates, WF2Q+ does not require reordering of buffered packets. The scheduling order is fully determined upon enqueue. Therefore, WF2Q+ does not rely on the update capability of $c.rank$ in UIFO, and it serves as a natural instance of UIFO's compatibility with PIFO/PIEO-model static-ordering algorithms. This further demonstrates that, while UIFO supports dynamic reordering semantics, it can seamlessly accommodate classical static ordering algorithms.
\subsection{Deficit Round Robin}
Deficit Round-Robin (DRR)\cite{drr} achieves fair service across multiple flows by performing round-robin scheduling among them, ensuring that each flow receives its fair share of service. DRR is a work-conserving scheduling algorithm, and can be viewed as a static-ordering algorithm. In UIFO, we map each flow to a class and element $e$ is modeled as an intra-flow data packet. Packets within a class are scheduled in FIFO order, while classes are scheduled in a round-robin fashion:
\begin{verbatim}
enqueue(e):{
    c = Class(flow)
    if(c not in class_list):
        c.rank = size(class_list) + 1
    else:
        c.rank = c.rank
    class_list.enqueue(c) # update c.rank
    e.rank = 1
    c.element_list.enqueue(e) 
}
dequeue():{
    c = class_list.head
    c.deficit_counter += quantum[c]
    while(c.element_list not empty 
          & (c.deficit_counter 
              >= size(c.element_list.head))):
          e = c.element_list.dequeue()
          c.deficit_counter -= size(e)
    if(c.element_list empty):
        c.deficit_counter = 0
        class_list.dequeue()
    else:
        c.rank += size(class_list) 
        class_list.enqueue(c) # update c.rank
}
\end{verbatim}
Upon enqueue, if the flow to which the arriving packet $e$ belongs is empty—i.e., the corresponding class does not yet exist in \textit{class\_list}—the scheduler inserts this class at the tail of \textit{class\_list}. Otherwise, the relative ordering among classes is preserved, and the packet is simply inserted into the \textit{element\_list} of its associated class. 
During dequeue, the scheduler selects the class at the head of \textit{class\_list}. If its \textit{element\_list} is non-empty and the current deficit is sufficient to transmit the head packet, the packet is dequeued and the deficit is updated accordingly. If the \textit{element\_list} becomes empty, the corresponding class is removed from \textit{class\_list}. If the deficit is insufficient, the priority of the class is updated to the lowest value, causing it to move to the tail of \textit{class\_list} to wait for the next scheduling round.
\\Under this mapping, UIFO accurately reproduces the scheduling semantics of DRR: flows are served in a round-robin manner, and whether a flow is scheduled depends on its deficit state. Since DRR does not rely on runtime state updates to reorder already-buffered packets, it does not require dynamic updates of $c.rank$ in UIFO. Instead, DRR is naturally supported as a static-ordering algorithm that operates at the flow granularity.
\subsection{Other examples}
For strict-priority algorithms that require only a single-level scheduling model, UIFO can express them by segmenting packet priorities into classes or by using the full priority field as the class classification criterion. Examples include EDF~\cite{edf}, LSTF~\cite{lstf}, SJF~\cite{sjf}, and SRTF~\cite{srtf}. Moreover, to address starvation issues inherent in strict-priority scheduling, UIFO can periodically update the ranks of low-priority classes to grant them service opportunities, thereby mitigating starvation without altering the overall scheduling framework.
\\UIFO supports scheduling over both class and element objects, and therefore is structurally well suited to express hierarchical scheduling algorithms. Since the \textit{element\_list} at the second level is functionally equivalent to a PIFO, it can be generalized into a scheduling tree. In this design, the first-level scheduler selects among different scheduling trees, while finer-grained packet-level or sub-class-level scheduling is performed within each tree. By adjusting the scheduling order of classes, the scheduler can control the service order among different scheduling trees, thereby enabling multi-level hierarchical scheduling policies.
\\The above examples demonstrate that UIFO can uniformly express scheduling algorithms operating at the granularity of packets, flows, or queues, encompassing both work-conserving and non-work-conserving algorithms, as well as static ordering and dynamic reordering semantics. By enabling scheduling over sets of schedulable objects and supporting priority updates over these sets, UIFO significantly extends the expressiveness of programmable data-plane scheduling abstractions while remaining compatible with existing models such as PIFO and PIEO.
\section{Hardware implementation}
\begin{figure}[t]
	\centering
	\includegraphics[width=0.98\columnwidth]{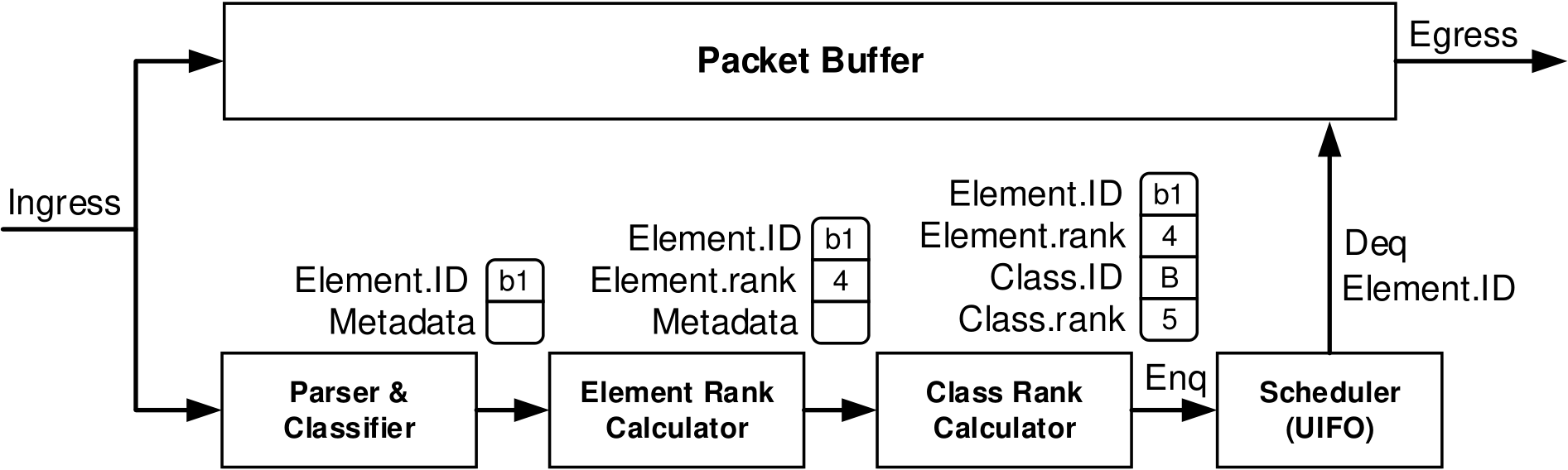}
	\caption{System Level Application For UIFO}
	\label{fig:3}
\end{figure}

\begin{figure}[htbp]
	\centering
	\includegraphics[width=0.98\columnwidth]{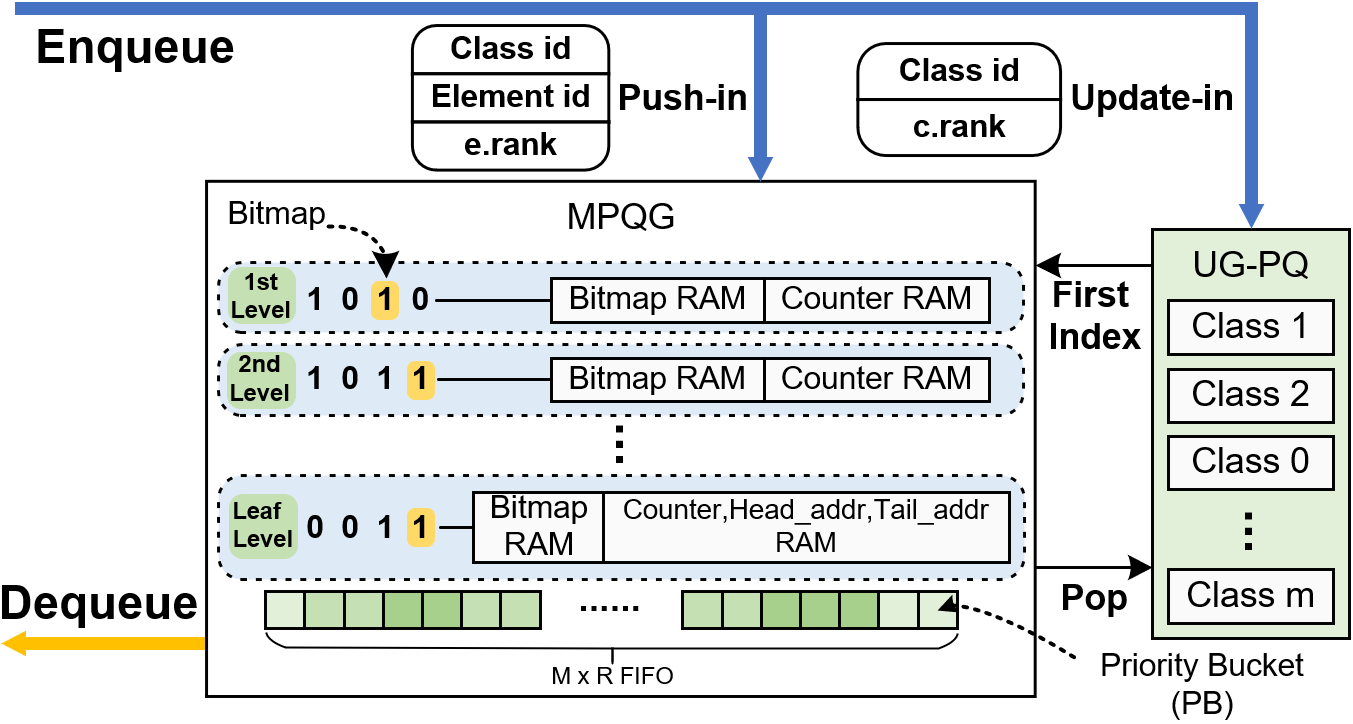}
	\caption{UIFO Hardware Implementation}
	\label{fig:4}
\end{figure}

In this section, we describe the hardware implementation architecture of UIFO. To deploy UIFO in a practical programmable data plane, it is necessary to clarify how UIFO is integrated at the hardware system level. Fig.~\ref{fig:3} illustrates the system-level integration of UIFO within the scheduling stage of a programmable data plane. It is important to emphasize that we intentionally unifies the classification results produced by the packet classifier with the notion of \textit{Class} in UIFO under a single abstraction. The classifier maps packets to logical categories—such as flows, queues, or priority groups—based on packet header fields and relevant state. Once classification is completed, subsequent scheduling decisions naturally operate at the granularity defined by this classification. From a semantic perspective, the classification outcome itself therefore defines the schedulable object. In UIFO, the scheduler directly treats the classification result as a class and performs scheduling and update operations accordingly. This paper does not incorporate the classification logic itself into the scheduling model; instead, it unifies the representation of classification outcomes and schedulable objects. This design choice improves the consistency between scheduling semantics and system implementation. UIFO assumes that classification results are provided externally and performs scheduling solely based on these results.
\\Packets first pass through the parser~\cite{parser} and classifier, where the \textit{Element ID} and relevant metadata are extracted. Subsequently, the element rank calculator and the class rank calculator compute $e.rank$ and $c.rank$, respectively. Packets carrying these attributes then enter the UIFO scheduling module. Upon dequeue, the scheduler retrieves the corresponding packet from the packet buffer using the \textit{Element ID} and transmits it.
\\The implementation of UIFO must simultaneously satisfy three key objectives: (1) supporting dynamic updates to class-level priorities; (2) enabling precise indexing and scheduling from \textit{class} to \textit{element\_list}; and (3) sustaining line-rate processing in a programmable data plane. To this end, we adopt a two-level, heterogeneous priority-queue architecture in hardware, with separate priority queues dedicated to inter-class scheduling and intra-class scheduling, respectively.
\\Since the first-level scheduling in UIFO operates on class objects and must support dynamic updates to $c.rank$, the corresponding priority queue is required to support update operations. In hardware, an update operation can be decomposed into two steps: deleting and reinserting the element. Existing priority queues that support deletion are typically implemented using one-dimensional structures~\cite{task_queue,antiq,arxiv}, such as shift registers, systolic arrays, or hybrid designs~\cite{systolic}.
\begin{table*}[htbp]
\centering    
\caption{Different Model Configurations}    
\label{tab:2}    
\footnotesize     
\setlength{\tabcolsep}{4pt} 
\begin{tabular}{lcccccc}
	\toprule        
	Model & Class Count & Class Priority Count & Predicate Count & Logical Partition Count & Element Priority Count & Queue Capacity 
	\\
	\midrule        
	UIFO & 256 & 256 & —  & —  & 256 & 128–65536
	\\ 
	PIEO~\cite{pieo} & — & —  & 256 & — & 256 & 128–65536
	\\
	PIFO-BBQ~\cite{bbq} & — & — & — & 256 & 256 & 128–65536
	\\
	\bottomrule
\end{tabular}
\end{table*}
In our implementation, we directly adopt the priority queue proposed in prior work~\cite{arxiv} by others as the base module. Throughout this paper, we refer to this structure as UG-PQ (Update and Group-Sorted Priority Queue). UG-PQ is built on a one-dimensional hybrid architecture, enabling simultaneous element's priority updates during enqueue operations. We do not modify the internal structure of UG-PQ in this work; its design details are not part of our contributions. In UIFO, UG-PQ is used to maintain \textit{class\_list}, enabling ordering and updates across classes.
\\The second-level scheduler is responsible for packet-level ordering. Since each class corresponds to an independent \textit{element\_list}, the overall structure constitutes a Multi-Priority-Queue Group (MPQG) composed of multiple priority queues. The MPQG must support: (1) precise indexing of the corresponding priority queue based on \textit{Class ID}; (2) insertion and dequeuing within that queue according to $e.rank$; and (3) adoption of a multi-list shared-memory structure to improve resource utilization. The bitmap-tree-based priority queue with logical partitioning~\cite{bbq} satisfies these requirements. It indexes \textit{Element IDs} into the corresponding priority buckets (PB) based on \textit{Class ID} and \textit{e.rank}, and uses linked lists to store elements belonging to the same PB. Therefore, we integrate the UG-PQ with a bitmap-tree-based MPQG to realize UIFO. The overall hardware architecture is shown in Fig.~\ref{fig:4}.
\\Upon enqueue, the \textit{Class ID} and $c.rank$ are inserted into the UG-PQ, which performs class ordering and priority updates. For classes with identical $c.rank$ values, their relative order in the UG-PQ follows a FIFO discipline. If the value of $c.rank$ remains unchanged before and after the update, the relative position of the class in the UG-PQ is preserved. In parallel, the \textit{Element ID} is indexed by its \textit{Class ID} and $e.rank$ and inserted into the corresponding PB in the MPQG. During dequeue, the scheduler first obtains the head class from the UG-PQ, then uses the bitmap-tree in conjunction with the Hierarchical Find-First Set principle to locate the head element of the priority queue corresponding to that class. When the \textit{element\_list} of the head class becomes empty, the first-level bitmap-tree triggers a pop operation on the UG-PQ, thereby removing the class.
\\A single operation on the UG-PQ requires three clock cycles. We accordingly modify the bitmap-tree-based MPQG so that each MPQG operation also completes in three cycles, eliminating data hazards due to pipeline techniques. For the leaf-level PB in the bitmap-tree-based MPQG, we employ a shared singly linked-list implementation, which helps reduce SRAM overhead. 
In terms of capacity configuration, if UIFO supports $M$ classes and $R$ element priority levels with a total queue capacity of $N$, the MPQG conceptually contains $M \times R$ FIFO queues. Owing to the shared-storage design, the effective maximum capacity of each FIFO queue remains $N$, thereby improving storage utilization while preserving full functionality.
\section{Evaluation}
We implement the above hardware architecture in Verilog and evaluate it on both a Xilinx XCVU13P FPGA platform and a 28nm ASIC process. We implement three designs—UIFO, PIEO, and PIFO-BBQ with logical partitioning~\cite{bbq}—and conduct a comparative analysis of their resource overhead and performance. It is worth noting that different designs may use different types of on-chip memory in FPGA synthesis results, such as URAM, BRAM, or LUTRAM. To avoid bias caused by differences in memory types, we directly extract the actual SRAM usage from the RTL code to account for storage overhead across all designs. UG-PQ~\cite{arxiv} can trade off between area and critical path length by configuring different numbers of shift registers within the systolic unit, while keeping the queue capacity unchanged. In our implementation, each systolic unit contains only 2 shift registers to minimize the critical path, which corresponds to the maximum area configuration. For the MPQG, we follow the conclusions in~\cite{bbq} and select the parameter configuration that yields the best performance, setting the bitmap width to 4. The parameter configurations for the three models are summarized in Table~\ref{tab:2}. Under these configurations, UIFO can support either a PIEO with equivalent functionality or a PIFO with the same number of logical partitions.
\begin{figure*}[t]
	\centering
	\begin{subfigure}[t]{0.32\linewidth}
		\centering
		\includegraphics[width=\linewidth]{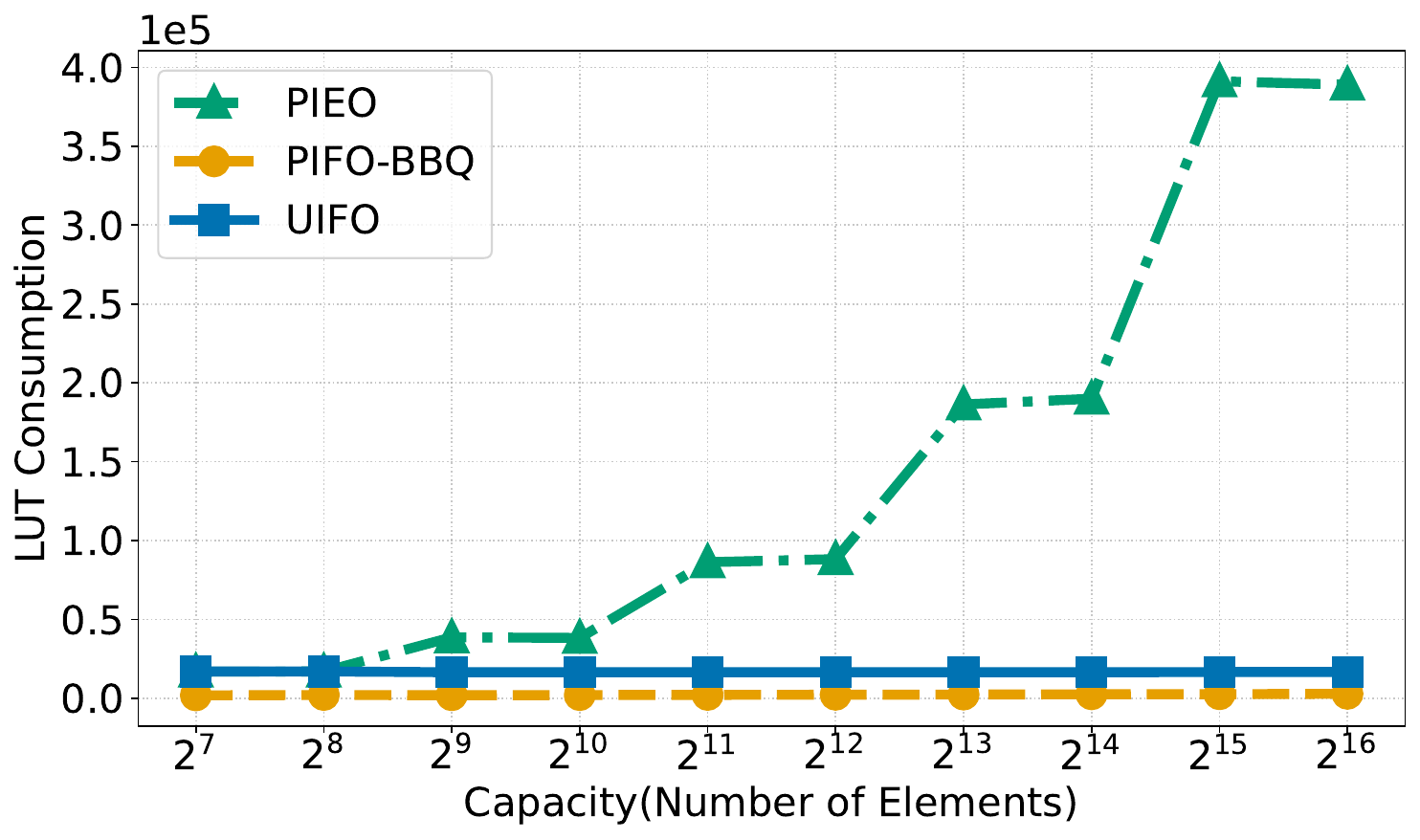}
		\caption{LUT consumption}
		\label{subfig:5a}
	\end{subfigure}\hfill
	\begin{subfigure}[t]{0.32\linewidth}
		\centering
		\includegraphics[width=\linewidth]{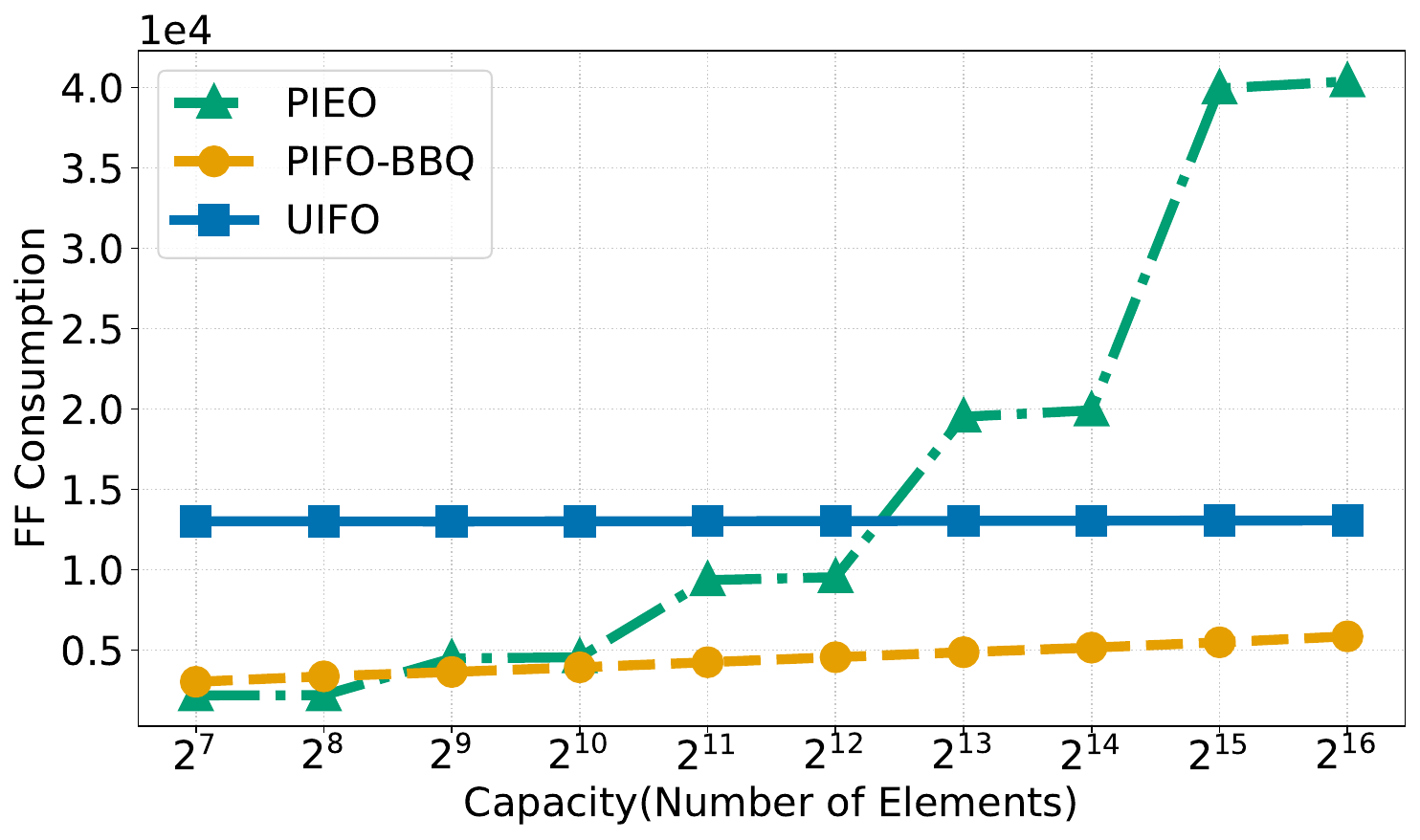}
		\caption{FF consumption}
		\label{subfig:5b}
	\end{subfigure}\hfill
	\begin{subfigure}[t]{0.32\linewidth}
		\centering
		\includegraphics[width=\linewidth]{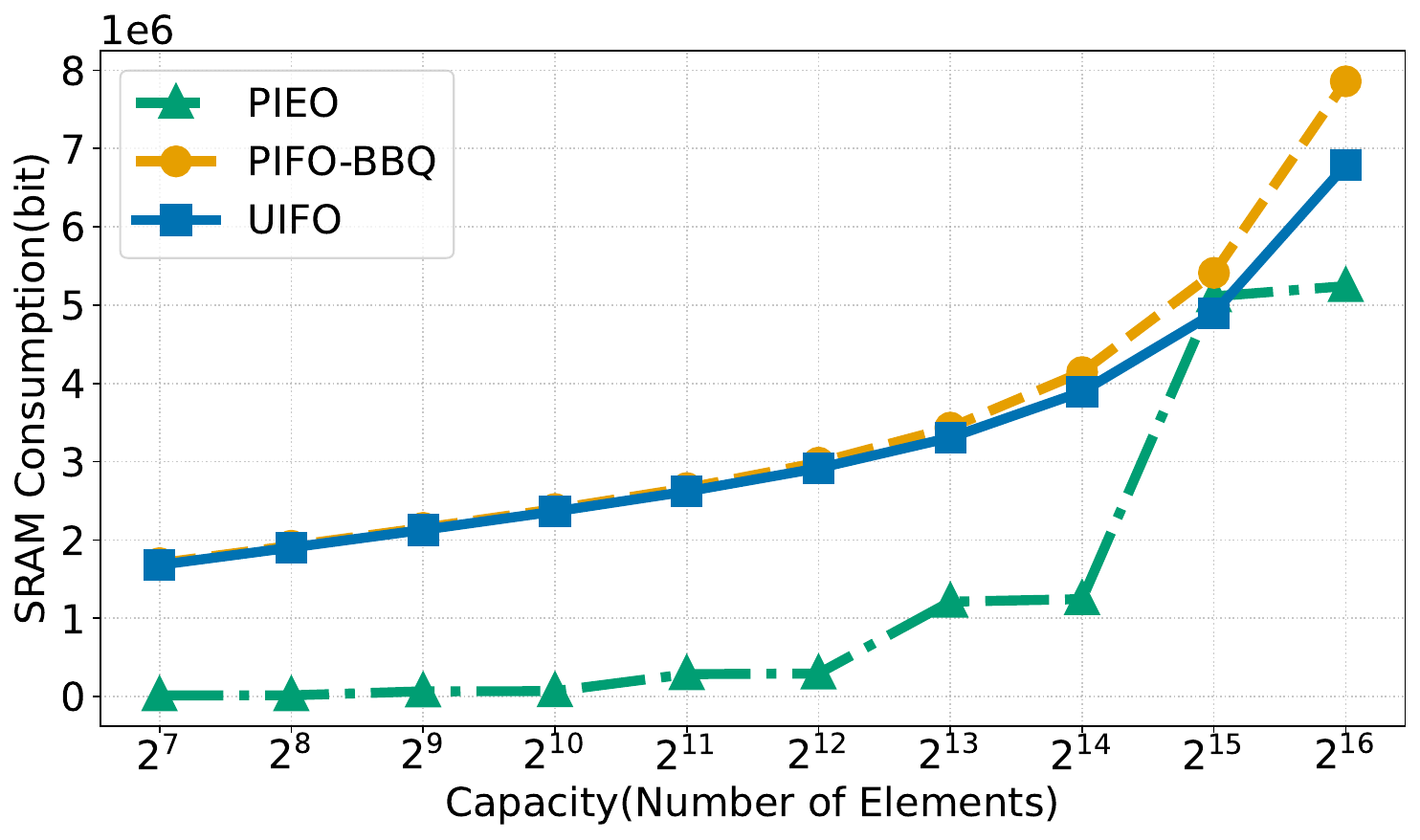}
		\caption{SRAM consumption}
		\label{subfig:5c}
	\end{subfigure}
	
	\caption{Comparison of Resource Overhead Among Three Models}
	\label{fig:5}
\end{figure*}
\subsection{Scalability}
In this section, we evaluate and compare the logic and storage resource overheads of the three scheduling model implementations. The comparison results are shown in Fig.\ref{fig:5}. In terms of LUT and FF usage, as the queue capacity increases, both PIFO-BBQ and UIFO remain largely unchanged. This is because packet elements are primarily stored in the MPQG implemented using a bitmap-tree structure; consequently, SRAM usage grows with queue capacity, while LUT and FF overheads remain relatively constant. However, UIFO incurs higher LUT and FF overhead than PIFO-BBQ: its LUT usage is approximately $6\times$--$9\times$ that of PIFO-BBQ, and its FF usage is about $2\times$--$4\times$. This increase stems from UIFO’s additional priority queue that supports update operations. Since changes in queue capacity do not affect the number of classes or class priorities, the overhead of this structure remains constant.
\\As queue capacity increases, UIFO and PIFO-BBQ exhibit similar SRAM scaling trends, with UIFO consistently using slightly less SRAM. This reduction is due to our use of a singly linked-list implementation for the MPQG. In practice, although PIFO-BBQ supports logical partitioning, it does not provide explicit control logic for selecting which logical partition to schedule. UIFO, by introducing explicit class-level scheduling and supporting dynamic class-priority updates, adds extra control logic overhead, which can be viewed as the cost of richer scheduling semantics.
\\Compared with PIEO, UIFO consistently incurs lower LUT overhead, and as queue capacity increases, PIEO’s FF usage eventually exceeds that of UIFO. This is because PIEO relies on comparator-based structures for element scheduling, whose logic overhead necessarily grows with queue capacity. By contrast, UIFO’s SRAM overhead mainly arises from the MPQG implementation, where bitmap and counter structures are maintained at each level; therefore, for small queue capacities, PIEO may use less SRAM. However, to guarantee constant-time operations, PIEO allocates twice the actual required SRAM capacity, causing its SRAM usage to grow faster than that of UIFO as queue capacity increases.
\\Finally, we note that when UIFO is configured in compatibility modes equivalent to PIEO or PIFO, its LUT and FF usage remains largely constant as queue capacity scales, and its SRAM growth trend matches that of PIFO-BBQ. This observation indicates that the generalized abstraction provided by UIFO does not introduce unnecessary control complexity.
\subsection{Scheduling rate}
\begin{figure}[tp]
	\centering
	\includegraphics[width=0.98\columnwidth]{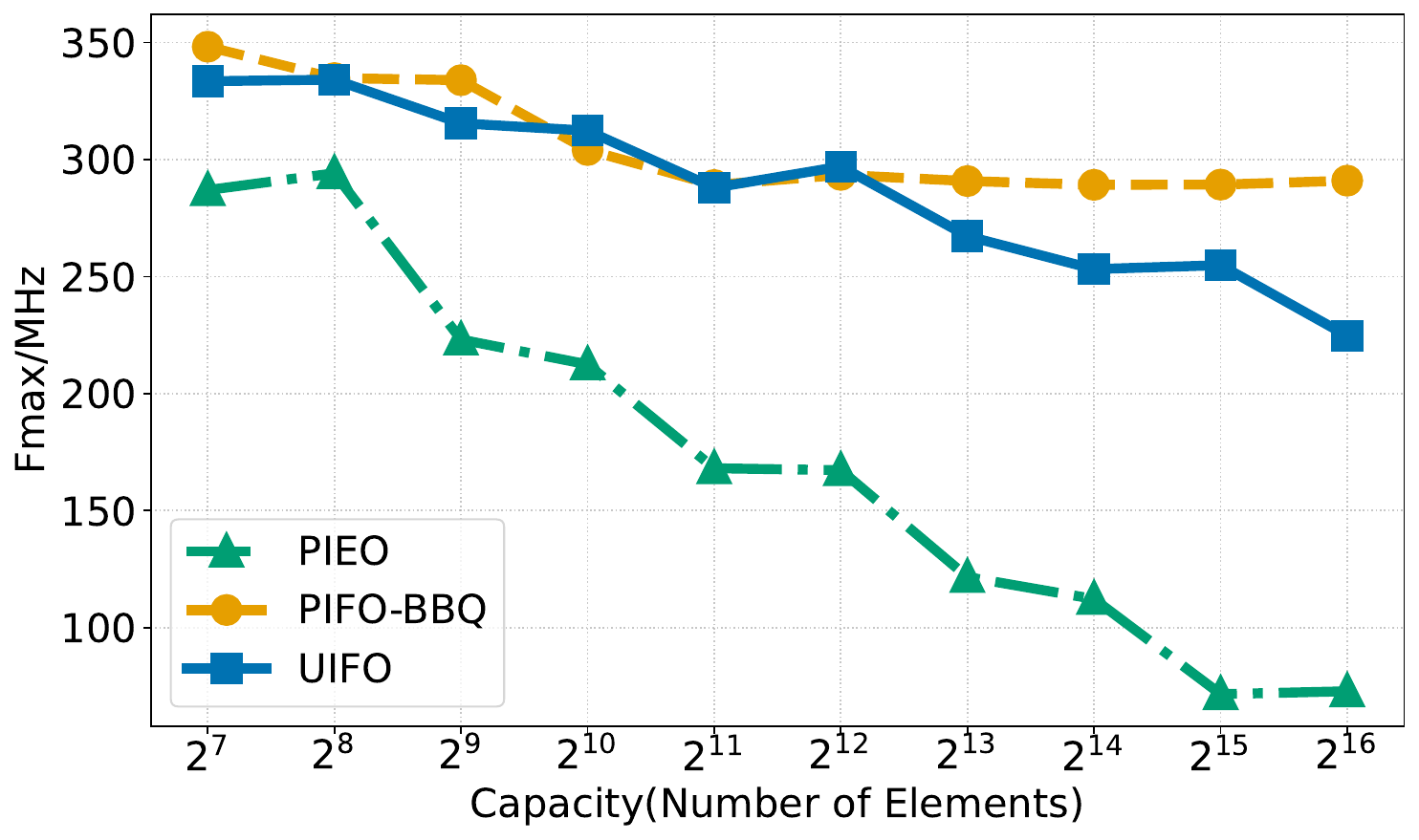}
	\caption{Maximum Operating Clock Frequency of Three Models}
	\label{fig:6}
\end{figure}

This section primarily evaluates the scheduling performance of UIFO in comparison with other schedulers. Fig.~\ref{fig:6} shows the maximum operating frequency of the three models as the queue capacity increases. As queue capacity grows, circuit complexity increases and FPGA routing becomes more challenging, which inevitably leads to a reduction in clock frequency.
We observe that PIFO-BBQ, implemented using a tree-based and pipelined structure, has its critical path in the Find-First-Set logic. As long as the bitmap width at each level remains unchanged, its timing performance is largely insensitive to queue capacity. In contrast, PIEO employs a compare-and-shift architecture, whose critical path delay increases rapidly with queue capacity. 
For UIFO, when the queue capacity is small, its operating frequency is comparable to that of PIFO-BBQ, reaching approximately 300\,MHz. As queue capacity increases, UIFO’s operating frequency decreases slightly due to higher routing complexity. Overall, however, UIFO maintains timing performance close to that of BBQ while significantly outperforming PIEO, despite providing substantially richer scheduling capabilities.
The priority queue in UIFO that supports update operations is still implemented using a compare-and-shift architecture\cite{systolic} and requires three cycles to complete a single operation. As a result, UIFO’s throughput is approximately one-third that of PIFO-BBQ. As shown in Fig.~\ref{fig:6}, the minimum throughput achieved by UIFO is 74\,Mpps. Under the configuration listed in Table~2, UIFO is able to sustain line-rate operation at 100\,Gbps for packets larger than 169\,bytes.
\subsection{ASIC Implementation}
In this section, we evaluate the area overhead and performance of UIFO under different configurations using a 28nm ASIC process. We synthesize two representative parameter configurations and queue capacities, with the results summarized in Table~\ref{tab:3}. The synthesis results show that SRAM accounts for the dominant portion of the overall area overhead. This is primarily because implementing the MPQG requires a large amount of on-chip storage to maintain element counters and corresponding bitmaps at each node of the indexing tree. As the number of supported element priority and the number of classes increase, the number of levels in the bitmap-tree grows, and the size of SRAM increases accordingly, thus becoming the dominant factor in area overhead.
Under the configuration with a queue capacity of 65536, UIFO exhibits a critical path delay of 2.12\,ns, enabling a throughput of 148.8\,Mpps. This result indicates that even under large-capacity configurations, UIFO is able to sustain 100Gbps line-rate operation requirements.
\begin{table}[tp]
	\centering
	\caption{ASIC Synthesis Results}
	\label{tab:3}
	\resizebox{\linewidth}{!}{
		\begin{tabular}{ccccccccc}
			\toprule
			Class & Class & Element & Element & Critical & \multicolumn{3}{c}{Area ($\text{mm}^2$)} \\
			Num & Rank & Num & Rank & Path (ns) & Logic & SRAM & Total \\
			\midrule
			64 & 64 & 4096 & 64 & 1.21 & 0.0140 & 0.0798 & 0.0938 \\
			256 & 256 & 65536 & 256 & 2.07 & 0.0553 & 1.5015 & 1.5568 \\
			4096 & 4096 & 65536 & 16 & 2.12 & 1.2639 & 1.4762 & 2.7401 \\
			\bottomrule
		\end{tabular}
	}
\end{table}
\section{Realated Work}
Existing hardware-realizable programmable packet scheduling models mainly fall into two categories: PIFO~\cite{pifo} and PIEO~\cite{pieo}. Since the introduction of the PIFO programmable packet scheduling model in 2016, a large body of follow-up work has focused on improving queue scalability and throughput, proposing various exact implementations such as SIMD PQ~\cite{simd-pq}, BMW-Tree~\cite{bmw}, and BBQ~\cite{bbq}. In addition, another line of work trades priority ordering accuracy for higher performance and scalability, including SP-PIFO~\cite{sp_pifo}, PCQ~\cite{pcq}, Gearbox~\cite{gearbox}, AIFO~\cite{aifo}, and Sifter~\cite{sifter}. However, all these efforts primarily target efficient implementations of the PIFO abstraction. From the perspective of supported scheduling algorithms, their semantics remain confined to packet-level ordering, making it difficult to express dynamic reordering algorithms that depend on runtime state updates.
\\To overcome the semantic limitations of PIFO, PIEO~\cite{pieo} was proposed in 2019. PIEO observes that PIFO cannot naturally express algorithms that simultaneously specify both transmission timing and service order, and introduces an \emph{eligibility predicate} to construct a subset of schedulable elements, from which the highest-priority packet is selected for scheduling. This design broadens the range of expressible scheduling policies. Subsequent approximate designs, such as PIPO~\cite{pipo} and CIPO~\cite{cipo} still fail to precisely implement dynamic reordering algorithms.
\\Prior work has largely concentrated on improving the scalability of PIFO and PIEO, whereas UIFO addresses their fundamental semantic limitations. By rethinking programmable scheduling from an abstraction perspective, UIFO strictly preserves compatibility with and generalizes existing PIFO and PIEO models, while enabling expressive and efficient support for dynamic ordering algorithms.
\section{Discussion}
From a data-structure perspective, UIFO more closely resembles a dictionary abstraction~\cite{dictionary}. In particular, UIFO requires the uniqueness of classes within the scheduler, a property that aligns well with dictionary data structures: a dictionary maintains a one-to-one mapping between keys and values while enforcing key uniqueness, and inserting an already existing key results in updating its associated value rather than creating a new entry. Viewed at an abstract level, a priority queue that supports update operations can thus be regarded as an ordered dictionary. UIFO leverages this property to enable dynamic updates to the priorities of schedulable objects at the class level.
\\From an implementation standpoint, UIFO combines a first-level update-enabled priority queue with a second-level PIFO-BBQ structure based on logical partitioning, allowing the two-level scheduling semantics defined by UIFO to be realized precisely in hardware. This observation suggests that UIFO’s hardware realization can be built by reusing and composing existing data-structure designs.
\\Finally, UIFO’s class-oriented scheduling mechanism need be tightly integrated with programmable parsers and classifiers based on the RMT architecture~\cite{rmt}, so that classes can be dynamically constructed and prioritized according to packet attributes. We believe that this design not only enhances the expressiveness of the scheduling model itself, but also further expands the flexibility of programmable data planes in queue management and scheduling, providing a unified abstraction foundation for more sophisticated scheduling policies and network functions in the future.
\section{Conclusion}
This paper proposes UIFO (Update-In-First-Out), a new programmable packet scheduling model that introduces a two-level scheduling structure over classes and packets. By elevating scheduling decisions from packet-level ordering to set-level scheduling and enabling dynamic updates to class priorities, UIFO significantly extends the expressive power of programmable schedulers. In terms of scheduling semantics, UIFO is strictly compatible with and generalizes existing PIFO and PIEO models, while naturally supporting dynamic-ordering algorithms that depend on runtime state changes. We believe that UIFO provides a set-oriented scheduling abstraction, which not only facilitates the extension of flexibility in scheduling for programmable data planes, but also lays the foundation for realizing more complex and dynamic network scheduling policies under hardware constraints.

\begin{acks}
ChatGPT was used solely for language translation. The authors take full responsibility for the content of this manuscript.
\end{acks}

\bibliographystyle{ACM-Reference-Format}
\bibliography{reference}

\end{document}